\documentclass[%
 aps,
 jmp,%
 amsmath,amssymb,
preprint,%
]{revtex4-1}

\usepackage{graphicx}
\usepackage{dcolumn}
\usepackage{bm}
\usepackage{epstopdf}
\usepackage{amssymb}
\usepackage{color}
\usepackage[colorlinks=true, letterpaper=true, pdfstartview=FitV, linkcolor=blue, citecolor=blue, urlcolor=blue]{hyperref}
\usepackage{lineno}

\begin{document}

\title{Large Out-of-Plane Piezoelectric Effect in Janus Ferromagnetic Semiconductor Monolayer of CrOFBr}
\author{Qiuyue Ma}
\author{Guochun Yang}
\author{Busheng Wang}
\author{Yong Liu}\email{yongliu@ysu.edu.cn}
\affiliation{State Key Laboratory of Metastable Materials Science and Technology \& Key Laboratory for Microstructural Material Physics of Hebei Province, School of Science, Yanshan University, Qinhuangdao 066004, China}

\begin{abstract}
The exploitation of piezoelectric ferromagnetism (PFM) in two-dimensional (2D) materials with large out-of-plane piezoelectric response is motivated not only by technological applications but also scientific interest. In this study, the CrONM monolayer family (N=F, Cl; M=Br, Cl) was investigated using first-principles calculations, revealing that the Janus CrOFBr monolayer exhibits intrinsic ferromagnetic semiconductor behavior along with a significant out-of-plane piezoelectric effect. The calculated out-of-plane piezoelectric strain coefficients \emph{d$_{31}$} and \emph{d$_{32}$} are up to 1.21 and 0.63 pm/V, respectively. These values are greater than those of the majority of 2D materials. Furthermore, our findings demonstrate that applying tensile strain can enhance the out-of-plane piezoelectric response, leading to a respective 27\% and 67\% augmentation in the piezoelectric strain coefficients  \emph{d$_{31}$} and \emph{d$_{32}$} compared to the unstrained configurations. This discovery holds great potential for propelling the field of nanoelectronics forward and facilitating the development of multifunctional semiconductor spintronic applications. Finally, by comparing \emph{d$_{31}$} and \emph{d$_{32}$} of the CrONM monolayer family (N=F, Cl; M=Br, Cl), we find that the magnitudes of \emph{d$_{31}$} and \emph{d$_{32}$} are correlated with the electronegativity difference between the M and N atoms. These findings provide valuable insights for the design of 2D piezoelectric materials with enhanced vertical piezoelectric responses.

\end{abstract}

\maketitle


\maketitle
\section{Introduction}
Since the discovery of graphene in 2004, two-dimensional (2D) materials have garnered significant attention theoretically and experimentally~\cite{1K-Science-2004,2S.-Rev.-2007,3K-Nature-2005,4E-Phys.-2007}. Their intrinsic conductivity and tunable properties make them highly promising prospects for the advancement of next-generation devices and technologies. Magnetically ordered 2D materials bring numerous possibilities for new device concepts and physical phenomena, especially 2D magnetic semiconductors, offering unique opportunities for the exploration of low-dimensional magnetism. The 2D ferromagnetic (FM) semiconductors are considered highly promising candidates for nanospintronic devices, combining the advantages of 2D, magnetism, and semiconductivity properties. Unfortunately, 2D intrinsic FM semiconductors are rare due to the incompatibility between semiconductivity and magnetism. Until 2017, the advent of two 2D intrinsic FM semiconductors (Cr$_2$Ge$_2$Te$_6$~\cite{C-Nature-2017} and CrI$_3$~\cite{B-Nature-2017}) opened up opportunities for achieving scalable spintronic architecture.

2D FM materials that incorporate other appealing electronic characteristics, such as topology, valley properties, and piezoelectricity, can enhance device capabilities and pave the way for innovative nanoscale devices. Topological features have been identified in 2D FM materials like Fe$_2$IX (X = Cl and Br)~\cite{S-Nanoscale-2021}, FeX$_2$ (X=Cl, Br, I)~\cite{X.-APL-2020}, and VSiGeN$_4$~\cite{S..-PRB-2022}. Valley properties have been observed in 2D FM 1\emph{T}-CrXY (X=S, Se, Te, Y=F, Cl, Br, I)~\cite{J.-PRB-2024}, VGe$_2$P$_4$~\cite{as.-PRB-2023}. In recent years, the piezoelectric properties of semiconductors have also attracted extensive research~\cite{29S-Nano-2019,30F-Solid -2022,31S-Phys-2021,32J-Nanotechnology-2021}. The piezoelectric effect is the phenomenon of electric dipole moments induced in non-centrosymmetric materials by mechanical stress~\cite{S-J. Appl-1997}. Reports on 2D piezoelectric materials are increasingly emerging, both theoretically and experimentally. A number of 2D multifunctional piezoelectric materials have been predicted by first-principles calculations, including Janus group-III chalcogenide~\cite{Y-Appl-2017}, group IV monochalcogenides~\cite{R-Appl-2015}, and transition metal dichalchogenides (TMD)~\cite{L-ACS-2017}. Experimentally, 2H-MoS$_2$ has emerged as a prototypical 2D piezoelectric material, and its discovery has significantly advanced the understanding and exploration of piezoelectric properties in 2D materials~\cite{W-Nature-2014}. By integrating piezoelectricity and magnetism within a 2D material, known as 2D piezoelectric ferromagnetism (PFM), significant advancements have been achieved in 2D vanadium dihalides, such as VS$_2$, VSe$_2$, and Janus VSSe, which not only exhibit magnetic semiconductor behavior but also demonstrate remarkable piezoelectric response~\cite{J-Phys-2019}.

In 2D materials, centrosymmetric structures inherently lack intrinsic piezoelectricity, thus requiring the exploration of different approaches to elicit piezoelectric responses. The construction of a Janus structure is a new method that breaks the mirror symmetry. The broken out-of-plane symmetry in these materials allows for the observation of out-of-plane piezoelectricity, offering the potential to identify materials with large out-of-plane piezoelectric response. Specifically, in Janus MXY materials with M sandwiched between X and Y, the different electronegativities of X and Y elements result in an obvious spontaneous polarization~\cite{L-ACS-2017}. Thus, the Janus method provides a pathway to explore piezoelectric responses in 2D materials, and several related studies have already been conducted. Additionally, Janus MXY monolayers such as MoSSe have been successfully synthesized, indicating potential for future applications~\cite{1J-ACS-2017}. On the other hand, the strain-tuned piezoelectric response of MoS$_2$~\cite{N.-J-2017} and Janus TMD monolayer~\cite{Dimple-J-2018} has demonstrated that strain engineering provides an alternative solution for finding a significant out-of-plane piezoelectric response. So far, great advances have been made on 2D piezoelectric materials. However, there is a main issue of 2D piezoelectric materials, which is that the out-of-plane piezoelectricity in known 2D materials is absent or weak. For example, the monolayer of InCrTe$_2$ exhibits significant FM coupling and in-plane magnetic anisotropy. However, it has a weak out-of-plane piezoelectricity (\emph{d$_{31}$}=0.39 pm/V). The strong out-of-plane piezoelectric effect and its inverse effect are highly desirable for piezoelectric devices, which is compatible with the bottom/top gate technologies. Therefore, it is necessary to find further 2D magnetic materials with large out-of-plane piezoelectric properties for experimental studies and possible device applications.

Considering the charming piezoelectric nature of the Janus structure, in this work, we constructed the Janus CrOFBr monolayer structure, which can be obtained by replacing the F layer in the CrOF monolayer with Br atoms. Our calculations show that the Janus CrOFBr monolayer is dynamically, thermally, and mechanically stable. It is found that the Janus CrOFBr monolayer is a large gap intrinsic FM semiconductor. As a result of the broken horizontal mirror symmetry, the Janus CrOFBr monolayer possesses only out-of-plane piezoelectric response. The predicted out-of-plane piezoelectric coefficients \emph{d$_{31}$} and \emph{d$_{32}$} are 1.21 and 0.63 pm/V, respectively, which are higher than those of many 2D materials. Similar to Janus CrOFBr monolayer, the Janus CrOClBr and CrOFCl monolayers are also intrinsic FM semiconductor with large out-of-plane piezoelectric response. Moreover, we investigate the strain effects on physical properties of the Janus CrOFBr monolayer. The findings suggest that the application of biaxial strain does not induce phase transition. Additionally, it is found that tensile strain can improve \emph{d$_{31}$} and \emph{d$_{32}$}. The Janus CrOFBr monolayer may be a promising candidate for applications in spin electronic and piezoelectric devices.

\section{Methods}
The first-principle calculations are performed using the projector augmented wave (PAW) method within density functional theory (DFT)~\cite{P.-Phys. Rev.-1964,W-Phys. Rev.-1965} as implemented in the plane wave code Vienna ab initio Simulation Package (VASP)~\cite{G-J. Non-Cryst. Solids-1995,G. Kresse-Computational Materials Science-1996,G-Phys. Rev. B-1999}. The exchange and correlation effects are tackled by the generalized gradient approximation (GGA) formulation of Perdew-Burke-Ernzerhof (PBE)~\cite{J. P-Phys. Rev. Lett.-1996}. In order to account for the on-site Coulomb correlation of Cr-3\emph{d} electrons, the GGA+U scheme (U = 7.0 eV) is employed for the Cr-3\emph{d} electron~\cite{V. I-Phys. Rev. B-1993,V. I-Phys. Rev. B-1991}, which has been tested in the previous works~\cite{N.-J. Am. Chem. Soc.-2018,A. K-Phys. Chem. Chem. Phys.-2020}. The k-point grids of 15 $\times$ 13 $\times$ 1 is employed to sample the Brillouin zone for the unit cell. The plane-wave cut-off energy of 500 eV, the total energy convergence criterion of 10$^{-6}$ eV, and force convergence criteria of less than 0.01 eV/{\AA} on each atom are used to attain reliable results. A vacuum layer of at least 15 {\AA} along the z direction is added to avoid the interactions between the neighboring images. The phonon spectrum with a 3 $\times$ 3 $\times$ 1 supercell was calculated self-consistently with the use of the Phonopy code~\cite{S.-Rev. Mod. Phys.-2001}. Ab inito molecular dynamics (AIMD) simulations in the canonical (NVT) ensemble were performed for 3000 fs at 300 K with a
Nos\'{e}-Hoover thermostat~\cite{G-J Chem Phys-1992}. The elastic stiffness tensor \emph{C$_{ij}$} and piezoelectric stress tensor \emph{e$_{ij}$} are calculated by using the strain-stress relationship (SSR) and density functional perturbation theory (DFPT) method~\cite{A-Phys. Rev. B-2008}.  The 2D elastic coefficients $C_{ij}^{2D}$ and piezoelectric stress coefficients $e_{ij}^{2D}$ have been renormalized by $C_{ij}^{2D}$ = \emph{Lz}$C_{ij}^{3D}$ and $e_{ij}^{2D}$ = \emph{Lz}$e_{ij}^{3D}$, where \emph{Lz} is the length of unit cell along the \emph{z} direction.

\begin{figure}[t!hp]
\centerline{\includegraphics[width=0.9\textwidth]{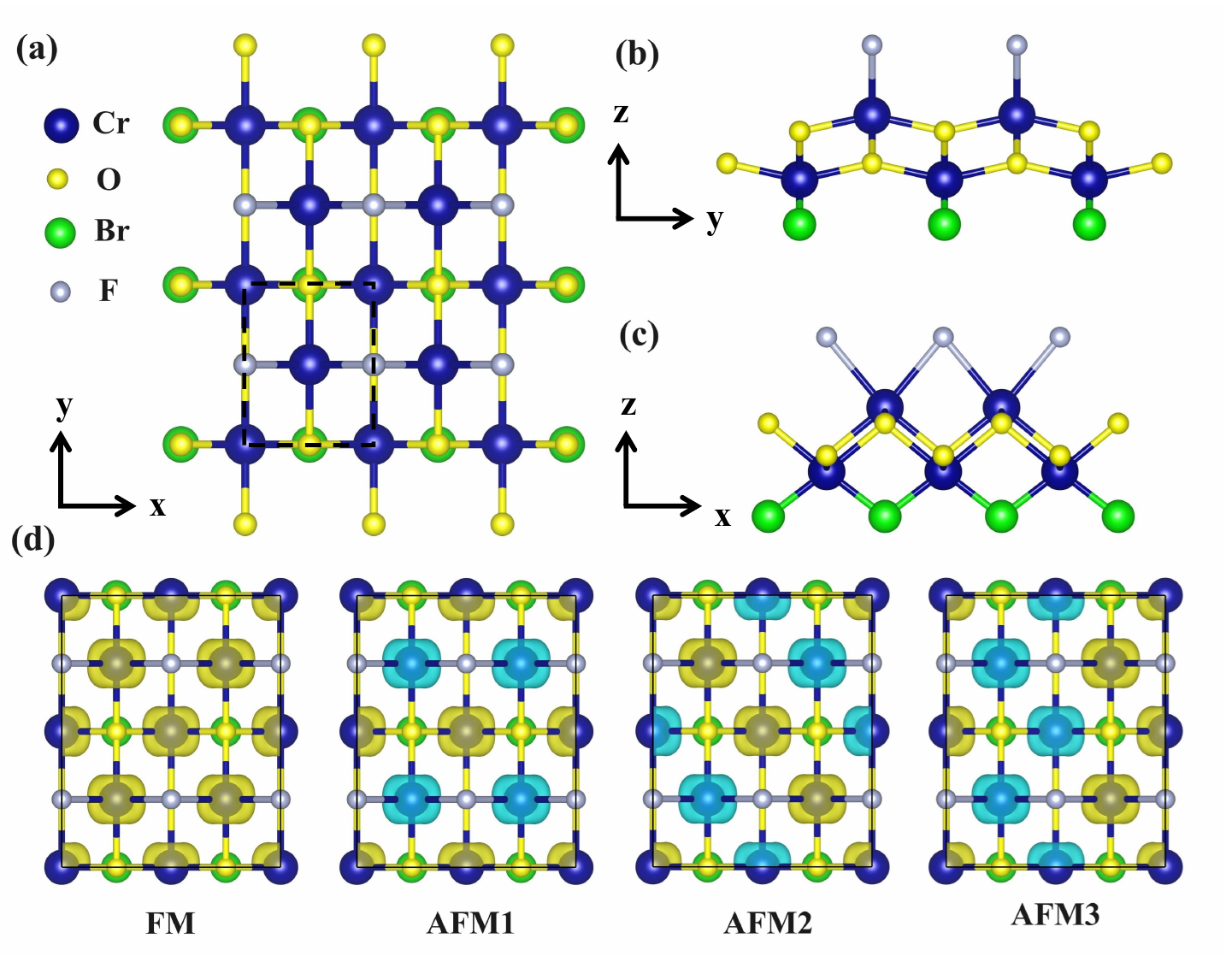}}
\caption{(a) Top and (b,c) side views of the Janus CrOFBr monolayer. (The area marked by black dotted line is the corresponding primitive cell.) (d) The spatial distribution of spin-polarized electron density for the Janus CrOFBr monolayer in One ferromagnetic configuration and three antiferromagnetic configurations.
\label{fig:stru0}}
\end{figure}


\section{Results and Discussion}
The top and side views of the Janus CrOFBr monolayer are shown in Fig.~\ref{fig:stru0} (a-c). It consists of a F-CrO-Br sandwich layer, which can be built by replacing one
of two F layers with Br atoms in a CrOF monolayer~\cite{C.-J. Phys-2021}. This gives rise to the formation of a typical Janus structure. The Janus CrOFBr monolayer has strongly distorted octahedron unit. Each Cr atom is 6-fold coordinated with four O, one F and one Br atom. The presence of broken vertical mirror symmetry in the CrOFBr monolayer leads to its space group being \emph{Pmm2} (No. 25), which is lower in symmetry compared to the \emph{Pmmn} space group of the CrOF monolayer (No. 59). For the CrOFBr monolayer, the disparity in atomic size and electronegativity between F and Br atoms leads to unequal bonding lengths and charge distributions in the Cr-F and Cr-Br bonds. This discrepancy generates an out-of-plane piezoelectric response and induces an electrostatic potential gradient. The optimized lattice parameters \emph{a} and \emph{b} are 3.202 and 3.977{\AA}, respectively. The bond lengths Cr-O, Cr-F, and Cr-Br of the Janus CrOFBr monolayer are found to be 2.04, 2.02, and 2.51 {\AA}. The magnetic moment of each primitive cell is calculated as 6 $\mu_B$, and the local magnetic moment per Cr is about 3.2 $\mu_B$. Similarly, Janus CrOFCl and CrOClBr monolayers can also be constructed from CrOF. Accordingly, the optimized lattice constants $a$, $b$, and bond lengths are summarized in Table~\ref{tablep1}. As illustrated in Fig.~\ref{fig:stru0}(d), we considered the FM and three antiferromagnetic (AFM) configurations for the supercells of 2 $\times$ 2 $\times$ 1 to determine the preferred magnetic ground state for Janus CrOFCl, CrOFBr, and CrOClBr monolayers.  Then, we performed spin-polarized DFT calculations. The energies of different magnetic configurations (FM, AFM1, AFM2, and AFM3) in meV per formula unit (f.u.) are detailed in Table SI of the supplementary material. The positive energy differences confirm that the FM configuration is the ground state for all three materials.

\begin{table}[h]
\begin{ruledtabular}
\caption{Lattice constants a and b ({\AA}), bond lengths ({\AA}), and  band gap (eV) of Janus CrOFCl, CrOFBr, and CrOClBr monolayers.}\label{tablep1}
\begin{tabular}{lcccccc}
     &a & b & \emph{L$_{Cr-O}$} & \emph{L$_{Cr-N}$} & \emph{L$_{Cr-M}$}& gap  \\
   \hline

CrOFCl & 3.16 & 3.97 &2.05 &2.00  &2.35 &  2.72  \\

CrOFBr & 3.20 & 3.98 & 2.04  &2.02 &2.51 & 1.67  \\

CrOClBr & 3.31 & 3.98 & 2.08 &2.39 &2.52 & 1.78\\
\end{tabular}
\end{ruledtabular}
\end{table}

\begin{figure}[t!hp]
\centerline{\includegraphics[width=0.9\textwidth]{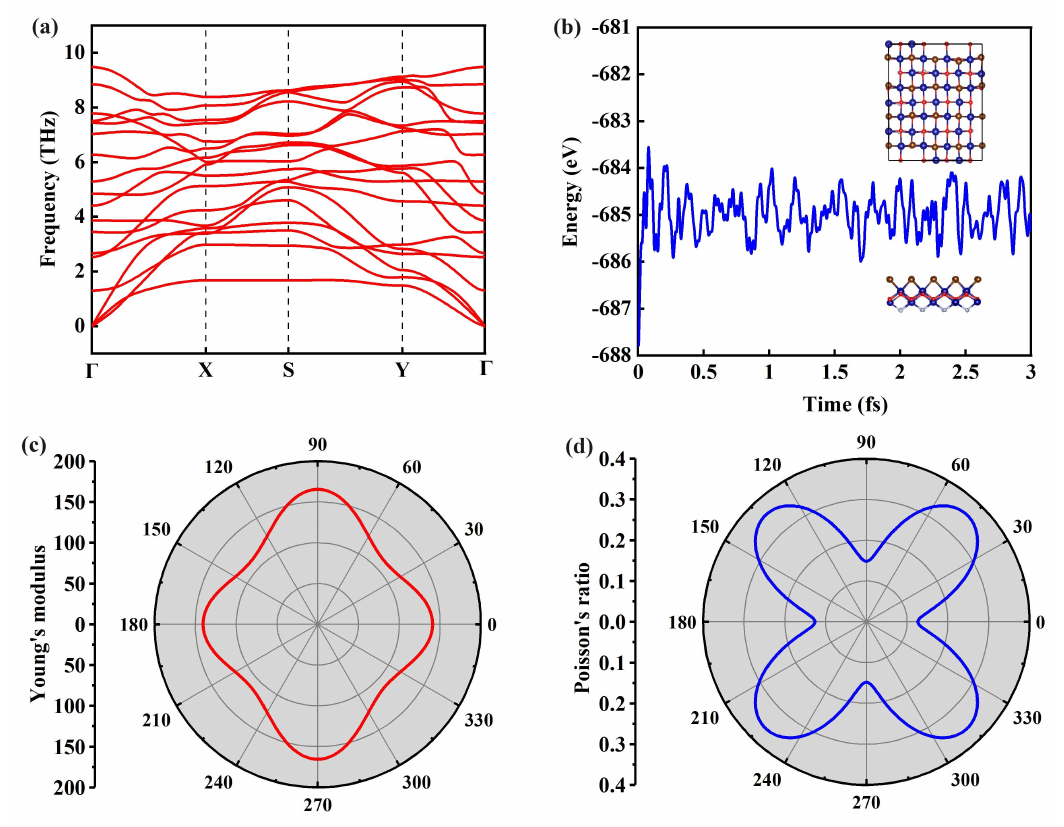}}
\caption{Determination of the stability for Janus CrOFBr monolayer, (a) The phonon spectra, (b) Total energy fluctuations with respect to AIMD simulation at 300 K. (c) The angular dependence of Poisson's ratio. (d) Young's modulus.
\label{fig:strun}}
\end{figure}

To verify the dynamical stability of the Janus CrOFBr monolayer, the phonon spectrum is calculated. As shown in Fig.~\ref{fig:strun}(a), the absence of imaginary-frequency modes in the phonon band dispersion suggests the Janus CrOFBr monolayer meets dynamic stability. The phonon spectrum presented in Fig. S1 of the supplementary material proves that Janus CrOFCl and CrOClBr monolayers are also dynamically stable. To further confirm the thermal stability, we performed AIMD simulations of the 4 $\times$ 4 $\times$ 1 supercells for 3000 fs at 300 K. Fig.~\ref{fig:strun}(b) shows the total energy fluctuations of the Janus CrOFBr monolayer as a function of the simulation time. The calculated results indicate no significant structural disturbance as the total energy fluctuates, confirming the thermodynamic stability of the Janus CrOFBr monolayer at room temperature. It is crucial to examine the practical applicability of the Janus CrOFBr monolayer. Therefore, we employed the SSR method to calculate the elastic stiffness tensor, allowing for the evaluation of the material's mechanical properties. The 2D elastic stiffness tensor (Voigt notation) with point group \emph{Pmm2} can be reduced into:

\begin{equation}
	\begin{pmatrix}
	 C_{11} & C_{12} & 0 \\
	C_{12} & C_{22} & 0 \\
	0 & 0 & C_{66}
	 \end{pmatrix}
\end{equation}

Our calculated elastic stiffness tensors \emph{C$_{11}$}, \emph{C$_{12}$}, \emph{C$_{22}$}, and \emph{C$_{66}$} are 143.31, 21.26, 168.62, and 41.51 N/m, respectively, which satisfy the Born-Huang criteria of mechanical stability (\emph{C$_{11}{>}$}0, \emph{C$_{22}{>}$}0, \emph{C$_{66}{>}$}0, \emph{C$_{11}$}-\emph{C$_{12}{>}$}0), thereby indicating the Janus CrOFBr monolayer is mechanically stable.
The Young's modulus Y$_{2D}$($\theta$) and Poisson's ratios $\nu_{2D}$($\theta$) can be calculated on the basis of the elastic stiffness tensors \emph{C$_{ij}$}. The in-plane Y$_{2D}$($\theta$) and $\nu_{2D}$($\theta$) can be calculated by the following two formulas~\cite{E-Phys. Rev. B-2012,E-Phys. Rev. B-2010}:

\begin{eqnarray}
{{Y_{2D}}(\theta) = \frac{{C_{11}}{C_{22}} - {C_{12}}^{2} }{{C_{11}}{m}^{4} + {C_{22}}{n}^{4} + (B - 2C_{12}){m}^{2}{n}^{2}} }
\end{eqnarray}
\begin{eqnarray}
{{\nu_{2D}}(\theta) = \frac{(C_{11} + C_{22} - B){m}^{2}{n}^{2} - {C_{12}({m}^{4} + {n}^{4})} }{{C_{11}}{m}^{4} + {C_{22}}{n}^{4} + (B - 2C_{12}){m}^{2}{n}^{2}} }
\end{eqnarray}

Where the $\theta$ is the angle of the direction with the x direction as 0$^\circ$ and the y direction as 90$^\circ$, \emph{m} = sin($\theta$), \emph{n} = cos($\theta$), and \emph{B} = ({C$_{11}$}{C$_{22}$} - {C$_{12}$}$^2$)/{C$_{66}$}. The Y$_{2D}$($\theta$) and $\nu_{2D}$($\theta$) as a function of the angle $\theta$ are plotted in Fig.~\ref{fig:strun}(c) and (d). Due to the CrOFBr monolayer is an asymmetric structure, the  Y$_{2D}$ and $\nu_{2D}$ of Janus CrOFBr monolayer show anisotropy along the (100) and (010) directions. The Y$_{2D}$ and $\nu_{2D}$ are 140.63/165.47 N/m and 0.126/0.148 along the (100)/(010) direction. The softest direction is along the (110) direction, with its Y$_{2D}$ of 112.83 N/m. And the maximun value of $\nu_{2D}$ is 0.359 along the (110) direction.

\begin{figure}[t!hp]
\centerline{\includegraphics[width=0.9\textwidth]{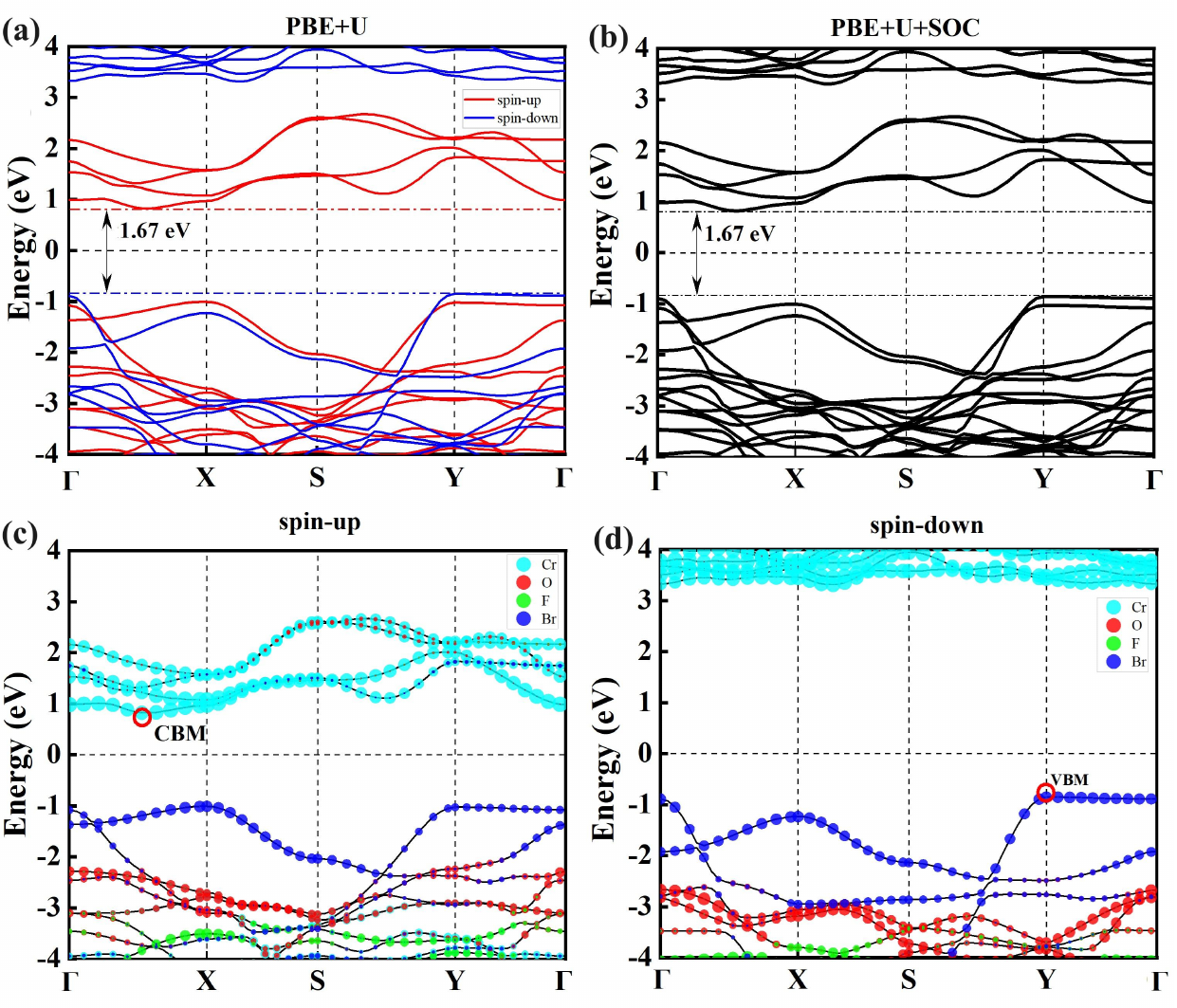}}
\caption{ The electronic band structures of Janus CrOFBr monolayer calculated by, (a) PBE + U, (b) PBE + U + SOC methods. The projected electronic band structures for (c) spin-up and (d) spin-down states of CrOFBr monolayer.
\label{fig:stru11}}
\end{figure}

As shown in Fig.~\ref{fig:stru11}(a) and (b), the band structures for Janus CrOFBr monolayer in ground magnetic state were investigated by using PBE + U and PBE + U + Spin-orbit coupling (SOC) methods. The PBE + U result shows that the Janus CrOFBr monolayer is an indirect gap semiconductor with a large band gap of 1.67 eV. The Janus CrOFBr monolayer exhibits a significant discrepancy of 2.49 eV between the conduction band minimum of its two spin channels, making it a promising candidate for spin polarized carrier injection and detection applications. Additionally, we employed the PBE + U + SOC method to calculate the band structure of the Janus CrOFBr monolayer. The results indicate that there is no significant difference in the electronic band structure (with a gap of 1.67 eV) after considering the SOC. This suggests that the electronic structure of the Janus CrOFBr monolayer has a negligible effect induced by SOC. To further explore the electronic properties of Janus CrOFBr monolayer, the projected electronic band structures as shown in Fig.~\ref{fig:stru11}(c) and (d). The valence-band maximum (VBM) of CrOFBr locates at point Y and the conduction-band minimum (CBM) at one point along the $\Gamma$-X path, respectively. For the purpose of comparing the electronic structural characteristics of CrOFBr with two other compounds, we have included the band structures of CrOFCl and CrOClBr in Fig. S2 of the supplementary material. It is evident that the band structures of all three compounds exhibit a remarkable similarity. Notably, CrOFCl stands out with the largest band gap, measuring 2.71 eV.

\begin{figure}[t!hp]
\centerline{\includegraphics[width=0.9\textwidth]{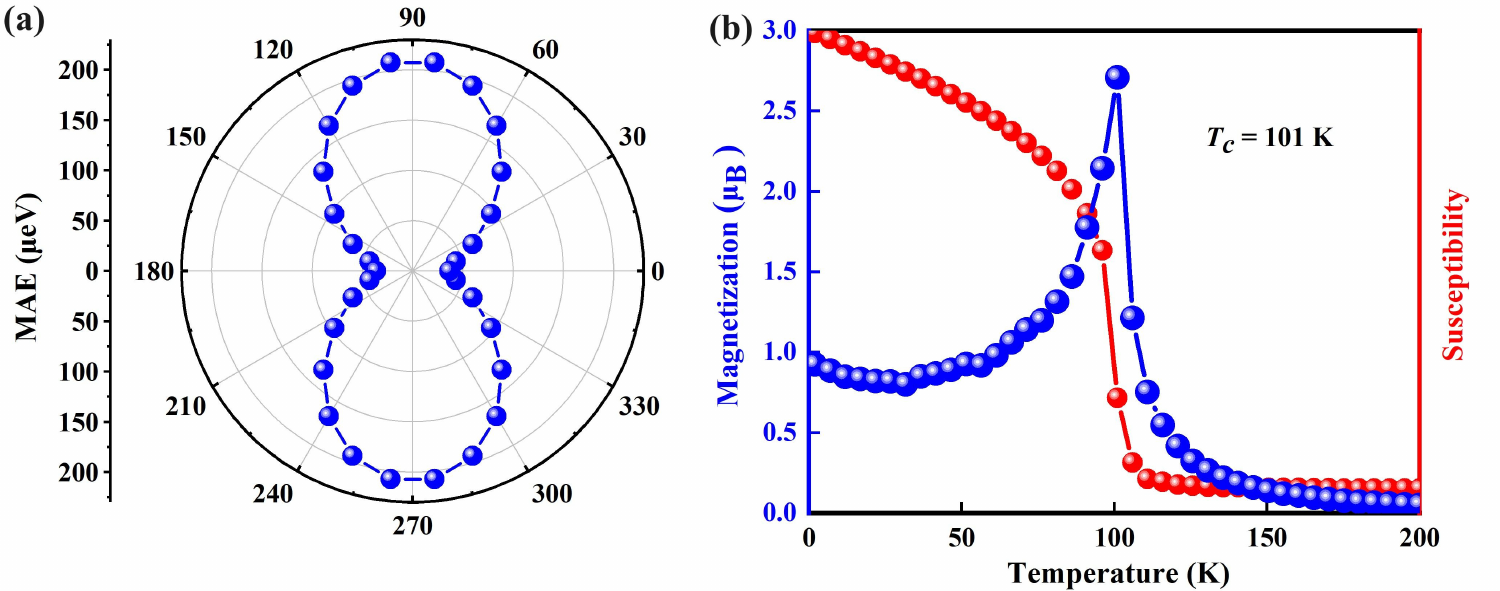}}
\caption{(a) Angular dependence of MAE of the Janus CrOFBr monolayer in the \emph{x}-\emph{y} plane. The relative energy of \emph{Mc} (001) is assigned as zero, representing the easy magnetic axis. (b) Magnetic moment and specific heat with respect to temperature for Janus CrOFBr monolayer.
\label{fig:stru2}}
\end{figure}

The magnetic anisotropy energy (MAE), derived from the spin-orbit coupling (SOC), plays a crucial role in understanding the stability and establishing the long-range magnetic order in low-dimensional magnets. The angular dependence of the MAE of the CrOFBr monolayer in the \emph{x}-\emph{y} plane is shown in Fig.~\ref{fig:stru2}(a). The relative energy of
\emph{M$_{C}$}[001] is set to zero. Consequently, The easy and hard magnetization axes correspond to the \emph{z}-axis and \emph{y}-axis, respectively, and the MAE of the CrOFBr monolayer reaches 0.21 meV per unit cell. To evaluate the potential practical applications of Janus CrOFBr monolayer in spintronic devices, we conducted additional Monte Carlo (MC) simulations with the Heisenberg model to calculate the variation trend of magnetism with temperature. The spin Hamiltonian can be considered as:
\begin{eqnarray}
{H =  - \sum\limits_{ < i,j > } {{J_{1}}{S_i}{S_j}}- \sum\limits_{ < k,l > } {{J_{2}}{S_k}{S_l}}- \sum\limits_{ < m,n > } {{J_{3}}{S_m}{S_n}}}
\end{eqnarray}

Here, the value of S is 3/2, which represents the spin of Cr atom. \emph{J$_1$}, \emph{J$_2$}, and \emph{J$_3$} represent the first-nearest, second-nearest, and third-nearest magnetic coupling parameters, respectively. Using the Heisenberg model Hamiltonian, the following energy equations can be written for different magnetic orders:

\begin{eqnarray}
{E(FM) = {E_0}- ({4J_1}+{2J_2}+{2J_3}){S^2} }
\end{eqnarray}
\begin{eqnarray}
{E(AFM1) = {E_0} - ({-4J_1}+{2J_2}+{2J_3}){S^2}}
\end{eqnarray}
\begin{eqnarray}
{E(AFM2) = {E_0} - ({-2J_2}-{2J_3}){S^2}}
\end{eqnarray}
\begin{eqnarray}
{E(AFM3) = {E_0} - ({-2J_2}+{2J_3}){S^2}}
\end{eqnarray}

The calculated the three exchange coupling parameters \emph{J$_1$}, \emph{J$_2$}, and \emph{J$_3$} are 1.52, 0.86 and 2.73 meV, respectively. Due to the positive values of \emph{J}, the FM configuration is the magnetic configuration of Janus CrOFBr monolayer. The variation of the magnetic moment and susceptibility as a function of temperature was considered through the MC simulation is illustrated in Fig.~\ref{fig:stru2}(b). Corresponding to the peak position of the specific heat (C$_V$), which defined by C$_V$=($ \langle  E^{2}  \rangle$)- $\langle  E  \rangle^{2}$)/\emph{K$_B$}$T^{2}$, and the local magnetic moments on Cr atoms begin to drop sharply, the calculated \emph{T}$_c$ of Janus CrOFBr monolayer is 101 K, which is higher than the \emph{T}$_c$ of Cr$_2$Ge$_2$Te$_6$ bilayer (30 K)~\cite{C-Nature-2017} and CrI$_3$ monolayer (45 K) ~\cite{B-Nature-2017}. Furthermore, the Janus CrOFCl and CrOClBr monolayers also possess out-plane magnetic anisotropy based on MAE, and their \emph{T}$_c$ values reach 110 K and 104 K, respectively.

The piezoelectric effect refers to the phenomenon in non-centrosymmetric materials where the application of strain or stress causes a change in charge distribution, leading to electric dipole moments and producing electricity. This phenomenon arises from the lack of inversion symmetry in the crystal structure of piezoelectric materials. The CrOFBr monolayer is a non-centrosymmetric structure, which means that it possesses a piezoelectric effect. Simultaneously, we calculated the average potential energy distribution along the z-axis, as depicted in Fig. S3 of the supplementary material. The results indicate a significant potential difference between the upper and lower surfaces of the 2D structure. Using the formula for the built-in electric field and considering the interatomic distance, we determined the intrinsic electric field strength of the Janus CrOFBr monolayer to be 1.03 eV/{\AA}. This value is notably high compared to those reported for other 2D piezoelectric materials, further suggesting the potential piezoelectricity in the Janus CrOFBr monolayer. Next, we specifically investigate the piezoelectric properties of the Janus CrOFBr monolayer. The piezoelectric effects of a material can be described by third-rank piezoelectric stress tensor \emph{e$_{ijk}$} and strain tensor \emph{d$_{ijk}$}, where the relaxed \emph{e$_{ijk}$} and \emph{d$_{ijk}$} are the sum of contributions from both ions and electrons:

\begin{eqnarray}
{\emph{e$_{ijk}$} = \frac{\partial{P_{i}}}{\partial{\varepsilon_{jk}}} = e_{ijk}^{elc} + e_{ijk}^{ion}}
\end{eqnarray}
and
\begin{eqnarray}
{\emph{d$_{ijk}$} = \frac{\partial{P_{i}}}{\partial{\sigma_{jk}}} = d_{ijk}^{elc} + d_{ijk}^{ion}}
\end{eqnarray}

The relationship between the \emph{e$_{ijk}$} and \emph{d$_{ijk}$} can be expressed as follows:
\begin{eqnarray}
{\emph{e$_{ijk}$} = \frac{\partial{P_{i}}}{\partial{\varepsilon_{jk}}} = \frac{\partial{P_{i}}}{\partial{\sigma_{mn}}}.\frac{\partial{\sigma_{mn}}}{\partial{\varepsilon_{jk}}} = d_{imn}C_{mnjk} }
\end{eqnarray}

where  \emph{P$_{i}$}, $\varepsilon$$_{jk}$ and $\sigma$$_{jk}$ represent the piezoelectric polarizations, strains,
and stresses, respectively. The C$_{mnjk}$ is elastic stiffness tensor. For 2D materials, only the in-plane strain and stress are taken into account~\cite{9M-ACS Nano-2015,10Y-Phys-2019,11S-Europhys-2020,12S-Comput.-2021,14Y-Appl-2017}. Using Voigt notation, the piezoelectric stress and piezoelectric strain tensors with \emph{Pmm2} point group symmetry can be expressed as:

\begin{equation}
	\begin{pmatrix}
	 0 & 0 & 0 \\
	0 & 0 & 0 \\
	e_{31} & e_{32} & 0
	 \end{pmatrix}
\end{equation}

\begin{equation}
	\begin{pmatrix}
	 0 & 0 & 0 \\
	0 & 0 & 0 \\
	d_{31} & d_{32} & 0
	 \end{pmatrix}
\end{equation}

When a uniaxial in-plane strain is imposed, the vertical piezoelectric polarization (\emph{d$_{31}$} $\neq$ 0 or \emph{d$_{32}$} $\neq$ 0) can be induced. However, by imposing biaxial in-plane strain, the superposed out-of-plane polarization will arise (\emph{d$_{31}$}$\neq$ 0 and \emph{d$_{32}$} $\neq$ 0). The \emph{d$_{31}$} and \emph{d$_{32}$} can be calculated by equations (11)-(13):

\begin{eqnarray}
{\emph{d$_{31}$} = \frac{{e_{31}}{C_{22}} - {e_{32}}{C_{12}} }{{C_{11}}{C_{22}} - {C_{12}}^{2}} }
\end{eqnarray}

\begin{eqnarray}
{\emph{d$_{32}$} = \frac{{e_{32}}{C_{11}} - {e_{31}}{C_{12}} }{{C_{11}}{C_{22}} - {C_{12}}^{2}} }
\end{eqnarray}

\begin{figure}[t!hp]
\centerline{\includegraphics[width=0.8\textwidth]{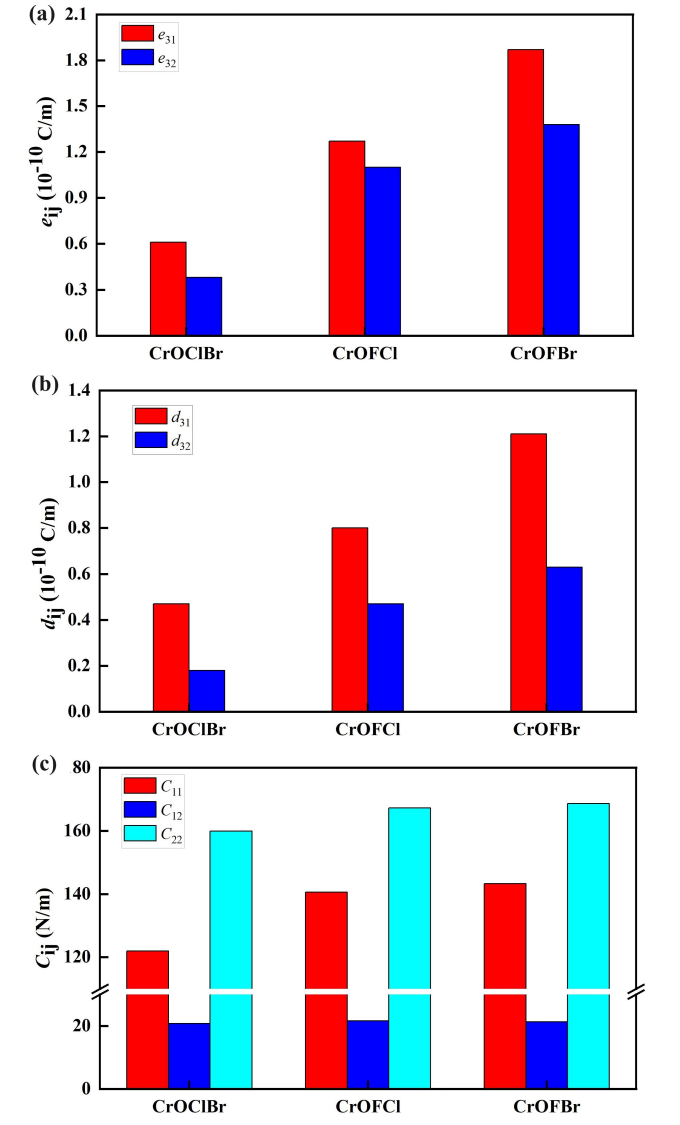}}
\caption{(a) elastic tensors \emph{C$_{11}$}, \emph{C$_{12}$}, and \emph{C$_{22}$}, (b) piezoelectric stress coefficients \emph{e$_{31}$} and \emph{e$_{32}$}, (c) piezoelectric strain coefficients \emph{d$_{31}$} and \emph{d$_{32}$} for Janus CrOFCl, CrOFBr, and CrOClBr monolayers.
\label{fig:stru8}}
\end{figure}

The \emph{e$_{31}$} and \emph{e$_{32}$} of Janus CrOFBr monolayer can be directly calculated by DFPT. The calculated \emph{e$_{31}$} (\emph{e$_{32}$}) is 1.87 $\times$ $10^{-10}$ C/m (1.38 $\times$ $10^{-10}$ C/m) with ionic part -0.40 $\times$ $10^{-10}$ C/m (-0.03 $\times$ $10^{-10}$ C/m) and electronic part 2.27 $\times$ $10^{-10}$ C/m (1.35 $\times$ $10^{-10}$ C/m). The contributions from both electrons and ions to \emph{e$_{31}$} and \emph{e$_{32}$} have opposite signs, with the electronic contribution dominating the piezoelectricity. According to equations (14)-(15), the calculated \emph{d$_{31}$} (\emph{d$_{32}$}) is 1.21 pm/V (0.63 pm/V). Specifically, the large out-of-plane piezoelectricity of Janus CrOFBr monolayer is significantly larger than those of most known 2D materials, like Janus group-III materials (0.46 pm/V)~\cite{Y-Appl-2017}, Janus transition metal dichalcogenide (TMD) monolayers (0.03 pm/V)~\cite{L-ACS-2017}, Janus BiTeI/SbTeI monolayer (0.37-0.66 pm/V)~\cite{40S-J. Appl-2020}, and oxygen funcitionalized MXenes (0.40-0.78 pm/V)~\cite{36J-Nano-2019}. The large out-of-plane piezoelectricity observed in this material provides broad prospects for the development and design of novel piezoelectric devices. Accordingly, we also calculate the piezoelectric properties of Janus CrOFCl and CrOClBr monolayers. The piezoelectric stress coefficients (\emph{e$_{31}$} and \emph{e$_{32}$}) and piezoelectric strain coefficients (\emph{d$_{31}$} and \emph{d$_{32}$}) of Janus CrOFCl and CrOClBr monolayers are plotted in Fig.~\ref{fig:stru8} (a) and (b), and the corresponding data are summarized in Table~\ref{tablep3}. It is found that \emph{d$_{31}$} and \emph{d$_{32}$} increase with the sequences of these monolayers (from CrOClBr to CrOFCl to CrOFBr). The electronegativity difference between Cl and Br atoms is 0.20, between F and Cl atoms is 0.82, and between F and Br atoms is 1.02. The results show that the \emph{d$_{31}$} and \emph{d$_{32}$} values in CrOFBr are the largest among these monolayers. This can be attributed to the largest electronegativity difference between F and Br. It is evident that the significant electronegativity difference between the two atoms plays a crucial role in enhancing the out-of-plane piezoelectric response.

\begin{table}[h]
\begin{ruledtabular}
\caption{Elastic constants \emph{C$_{ij}$} (N/m), and the piezoelectric coefficients \emph{e$_{ij}$} ($10^{-10}$ C/m) and \emph{d$_{ij}$} (pm/V) for Janus CrOClBr, CrOFCl, and CrOFBr monolayers.}\label{tablep3}
\begin{tabular}{lcccccccccccc}
     &\emph{C$_{11}$} & \emph{C$_{12}$} & \emph{C$_{22}$} & \emph{e$_{31}$}  & \emph{e$_{32}$} & \emph{d$_{31}$ }&\emph{d$_{32}$} \\
   \hline

CrOClBr & 121.94 & 20.78 & 159.88 & 0.61 & 0.38 & 0.47 & 0.18 \\

CrOFCl & 140.60 & 21.65 & 167.23 & 1.27 & 1.10 & 0.80 &  0.47\\

CrOFBr & 143.31 & 21.26 & 168.62 & 1.87 & 1.38 & 1.21 & 0.63\\
\end{tabular}
\end{ruledtabular}
\end{table}

\begin{figure}[t!hp]
\centerline{\includegraphics[width=0.9\textwidth]{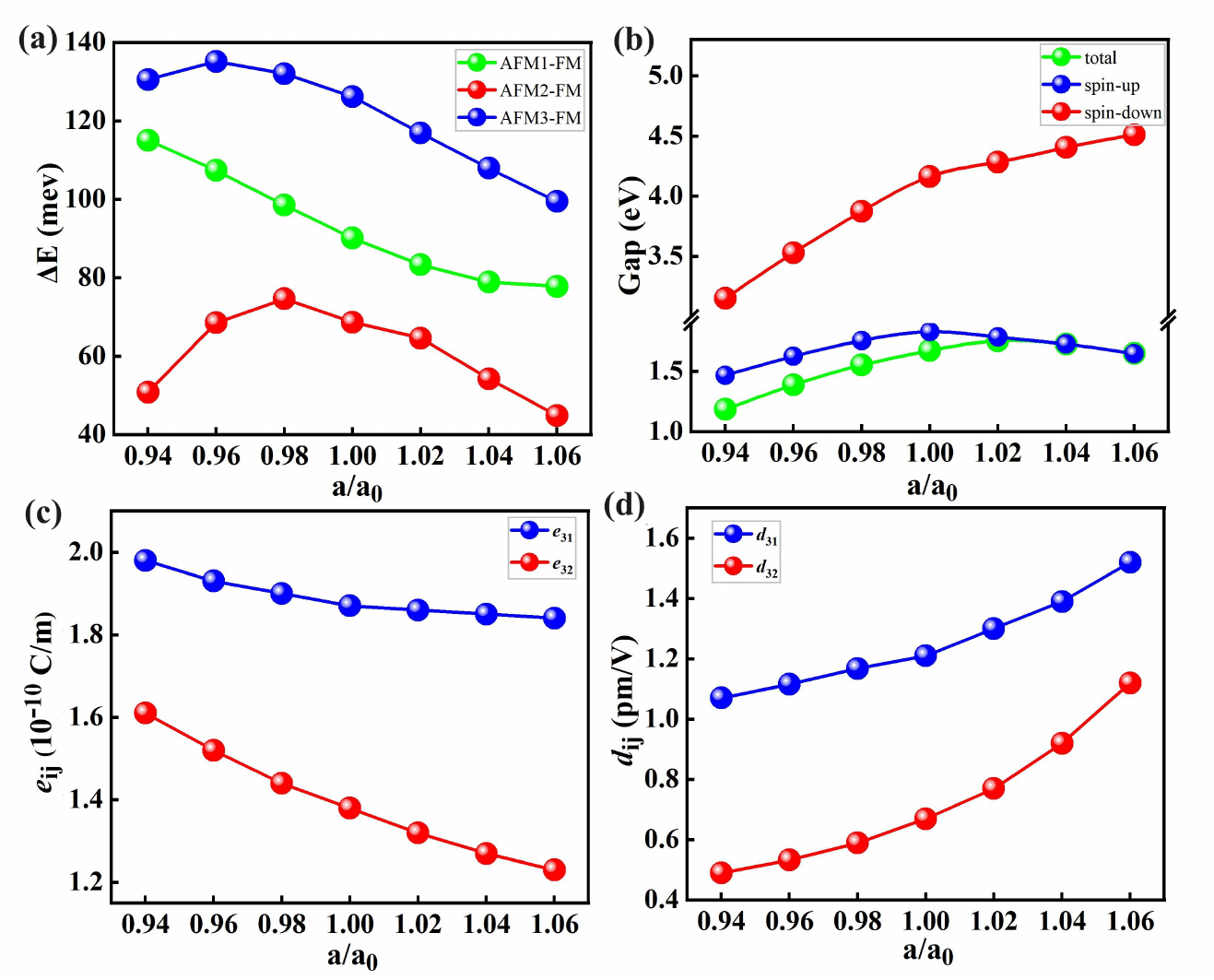}}
\caption{For Janus CrOFBr monolayer, (a) the energy differences of AFM1, AFM2, and AFM3 with respect to FM state, (b) the spin-up gap, the spin-down gap, and the total gap, (c) piezoelectric stress coefficients, (d) piezoelectric strain coefficients as a function of \emph{a}/\emph{a$_{0}$}.
\label{fig:struw}}
\end{figure}

Next, the influence of strain on the electronic structures and piezoelectric properties of the Janus CrOFBr monolayer was investigated. Studies have demonstrated that strain engineering can effectively modulate the electronic structures and piezoelectric characteristics of 2D materials ~\cite{Dimple-J-2018,51Dimple-J. Mater-2018,52S-J. Phys-2021}. Here, the \emph{a}/\emph{a$_{0}$} is used to simulate the biaxial strain, where \emph{a} and \emph{a$_{0}$} represent the strained and unstrained lattice constants, respectively. We investigated the effects of biaxial strain ranging from 0.94 to 1.06 on the properties of the Janus CrOFBr monolayer. Fig.~\ref{fig:struw}(a) illustrates the variation of the energy difference between the FM and three AFM configurations as a function of strain, all of which exhibit positive values. This indicates that the FM state always corresponds to the ground state. The energy band structures of the Janus CrOFBr monolayer, with the biaxial strain (\emph{a}/\emph{a$_{0}$}) ranging from 0.94 to 1.06, are depicted in Fig. S4 of the supplementary material. It indicates that the CrOFBr is always an indirect band gap semiconductor at different strains. At a strain ranging from 0.94 to 1.06, the conduction band minimum (CBM) is located at one point along the $\Gamma$-X path. At a compressive strain, the valence band maximum (VBM) of the Janus CrOFBr monolayer is located at Y point; however, the VBM moves from point Y to point X at a tensile strain of 1.02. The spin-up gap, the spin-down gap, and the total gap of Janus CrOFBr monolayer as a function of strain \emph{a}/\emph{a$_{0}$} are plotted in Fig.~\ref{fig:struw}(b). It is clearly seen that the strain ranging from 0.94 to 1.06 causes the conduction band and valence band of the spin-down channel to move away from the Fermi level, resulting in an increase in the spin-down gap. The total band gap undergoes an initial increase and then decreases within the strain range, reaching a maximum value of 1.76 eV at a strain of 1.02. The results show that the spin up and total gaps coincide at a strain of 1.02 to 1.06, which means that Janus CrOFBr monolayer exhibits half semiconductor character. Additionally, the piezoelectric stress coefficients (\emph{e$_{31}$} and \emph{e$_{32}$}) and piezoelectric strain coefficients (\emph{d$_{31}$} and \emph{d$_{32}$}) of Janus CrOFBr monolayer as a function of biaxial strain are shown in Fig.~\ref{fig:struw}(a). It is found that compressive strain can enhance \emph{e$_{31}$} (\emph{e$_{32}$}), and at 0.94 strain, \emph{e$_{31}$} (\emph{e$_{32}$}) improves to 1.98 (1.61) pm/V from the unstrained value of 1.87 (1.38) pm/V. On the other hand, tensile strain can decrease \emph{e$_{31}$} (\emph{e$_{32}$}), and at 1.06 strain, \emph{e$_{31}$} (\emph{e$_{32}$}) reduces to 1.84 (1.23) pm/V. Both the \emph{d$_{31}$} and \emph{d$_{32}$} coefficients of the Janus CrOFBr monolayer exhibit an increase as the strain varies from 0.94 to 1.06. The value of \emph{d$_{31}$} and \emph{d$_{32}$} reach a maximum of 1.52 and 1.12 pm/V at a strain of 1.06, which represents a 27\% and 67\% increase relative to the unstrained value. Meanwhile, we also calculated the effect of strain on the piezoelectric strain coefficients (\emph{d$_{31}$} and \emph{d$_{32}$}) of Janus CrOClBr and CrOFCl monolayers, as shown in Fig. S5 of the supplementary materials. It is clearly seen that the out-of-plane strain piezoelectric coefficients are also significantly enhanced as the strain value (\emph{a}/\emph{a$_{0}$}) increases.

\section{Conclusion}
In summary, we predicted that the Janus CrOFBr monolayer with space group \emph{Pmm2}, characterized by outstanding dynamic, thermal, and mechanical stabilities, demonstrates intrinsic FM semiconductor behavior based on first-principles calculations. Owing to structural symmetry breaking, the Janus CrOFBr monolayer only possesses out-of-plane piezoelectric response. The calculated out-of-plane piezoelectric coefficients \emph{d$_{31}$} and \emph{d$_{32}$} are 1.21 and 0.63 pm/V, respectively, significantly exceeding those found in the majority of known 2D materials. This significant out-of-plane piezoelectricity indicates that the monolayer has great potential for the operation and design of new piezoelectric devices. Additionally, the Janus CrOClBr and CrOFCl monolayers are also promising intrinsic FM semiconductor with large out-of-plane piezoelectric coefficients. It is clearly seen that the Janus CrOFBr monolayer prefers the FM ground state within a strain range from 0.94 to 1.06. Unexpectedly, the tensile strain can further enhance the out-of-plane piezoelectric strain coefficients \emph{d$_{31}$} and \emph{d$_{32}$}, which represent a 27\% and 67\% increase relative to the unstrained values. We believe that our study provides significant insights into the development of 2D piezoelectric materials characterized by a large vertical piezoelectric response and holds the potential to advance the field of nanoelectronics and enable the development of multifunctional semiconductor spintronic applications.

\section{Acknowledgments}

This work was supported by the Natural Science Foundation of China under Grants (No. 22372142), the Innovation Capability Improvement Project of Hebei province (No. 22567605H), the Natural Science Foundation of Hebei Province of China (No. B2021203030), the Science and Technology Project of Hebei Education Department (No. JZX2023020). The numerical calculations in
this paper have been done on the supercomputing system in the High Performance Computing Center of Yanshan University.



\end{document}